\documentclass[aps, prl, twocolumn, superscriptaddress, showpacs]{revtex4}
\usepackage{graphicx}
\usepackage{amsmath,amssymb}
\begin{document}

\title{Kondo effect in carbon nanotube quantum dots with spin-orbit coupling}
\author{Tie-Feng Fang}
\affiliation{Institute of Modern Physics, Chinese Academy of
Sciences, Lanzhou 730000, China}
\author{Wei Zuo}
\affiliation{Institute of Modern Physics, Chinese Academy of
Sciences, Lanzhou 730000, China}
\author{Hong-Gang Luo}
\affiliation{Center for Interdisciplinary Studies, Lanzhou
University, Lanzhou 730000, China} \affiliation{Key Laboratory for
Magnetism and Magnetic Materials of the Ministry of Education,
Lanzhou University, Lanzhou 730000, China} \affiliation{Institute of
Theoretical Physics, Chinese Academy of Sciences, Beijing 100080,
China}

\begin{abstract}
Motivated by recent experimental observation of spin-orbit coupling
in carbon nanotube quantum dots [F. Kuemmeth \textsl{et al.}, Nature
(London) {\bf 452}, 448 (2008)], we investigate in detail its
influence on the Kondo effect. The spin-orbit coupling intrinsically
lifts out the fourfold degeneracy of a single electron in the dot,
thereby breaking the $SU(4)$ symmetry and splitting the Kondo
resonance even at zero magnetic field. When the field is applied,
the Kondo resonance further splits and exhibits fine multipeak
structures resulting from the interplay of spin-orbit coupling and
Zeeman effect. Microscopic cotunneling process for each peak can be
uniquely identified. Finally, a purely orbital Kondo effect in the
two-electron regime is also predicted.
\end{abstract}
\maketitle

{\it Introduction.---}The Kondo effect is one of the most
fascinating and extensively studied subjects in condensed matter
physics \cite{book1}. Since its first experimental observation in
semiconductor quantum dots (QDs) in 1998 \cite{Goldhaber-Gordon},
after 10 years of theoretical predictions \cite{Ng1988}, various
aspects of this effect have been explored in virtue of the
tunability of relevant parameters in the QDs. The Kondo physics was
further enriched by fabricating carbon nanotube (CNT) QDs
\cite{Nygard, Buitelaar, Jarillo-Herrero1, Jarillo-Herrero2,
Makarovski1, Makarovski2} where additional orbital degree of freedom
originating from two electronic subbands can play a role of
pseudospin. In CNT QDs, spin-orbit coupling is widely believed to be
weak and two sets of spin-degenerate orbits are expected to yield a
fourfold-degenerate energy spectrum which possesses an $SU(4)$
symmetry. Consecutive filling of these orbits forms four-electron
shells \cite{Buitelaar, Jarillo-Herrero1, Jarillo-Herrero2,
Makarovski1, Makarovski2, Liang}. In each shell, the $SU(4)$ Kondo
effect was observed in the valleys with one, two, and three
electrons \cite{Jarillo-Herrero1, Jarillo-Herrero2, Makarovski1,
Makarovski2}. Theoretically, the $SU(4)$ Kondo effect in CNT QDs has
also been extensively studied \cite{Choi, Lim, Busser, Anders}

However, recent theories \cite{Huertas-Hernando2006} suggest that
spin-orbit interaction can be significant in CNTs due to their
curvature and cylindrical topology. More recently, transport
spectroscopy measurements on ultra-clean CNT QDs by Kuemmeth
\textsl{et al.}\;\cite{Kuemmeth} demonstrate that the spin and
orbital motion of electrons are strongly coupled, thereby breaking
the $SU(4)$ symmetry of electronic states in such systems. This
motivates us to reconsider the Kondo effect in CNT QDs by explicitly
taking account of the spin-orbit coupling since this
symmetry-breaking perturbation at the fixed point must break the
$SU(4)$ Kondo effect studied previously. An important consequence is
that even at zero magnetic field the Kondo effect manifests as split
resonant peaks in the differential conductance. At finite fields,
these peaks further split into much complicated subpeaks, reflecting
the entangled interplay of spin and orbital degrees of freedom.
Concerning all microscopic cotunneling events involving spin and/or
orbit flip, these fine multipeak structures can be uniquely
identified. Moreover, the spin-orbit coupling also determines the
filling order in the two-electron ($2e$) ground state
\cite{Kuemmeth}, producing a purely orbital Kondo effect different
from that observed by Jarillo-Herrero \textsl{et
al.}\;\cite{Jarillo-Herrero1}.

{\it Model Hamiltonian and QD Green's function.---}We model a CNT QD
coupled to source and drain leads by the Anderson Hamiltonian $H =
H_d + H_L + H_T$, where $H_{d} =
\sum_{m}\varepsilon_{m}d_{m}^{\dagger}d_{m}+\frac{U}{2} \sum_{m\neq
m^{\prime}}\hat{n}_{m}\hat{n}_{m^{\prime}}$, $H_{L} =
\sum_{k,m,\alpha}\varepsilon_{k}C_{km\alpha}^{\dagger
}C_{km\alpha}$, $H_{T}
=\sum_{k,m,\alpha}V_{\alpha}d_{m}^{\dagger}C_{km\alpha
}+\textrm{H.c.}.$ $H_d$ models the isolated CNT QD with $\hat{n}_{m}
= d_{m}^{\dagger}d_{m}$. $d^\dagger_m$ ($d_m$) creates (annihilates)
an $m=\{\sigma,\,\lambda\}$ configuration electron in the dot, where
$\sigma,\lambda=\pm$ are the spin and orbital quantum numbers,
respectively. $U$ denotes the on-site Coulomb repulsion, and
$\varepsilon_{m}$ is the single-particle energy. $H_L$ models the
two leads ($\alpha=L,R$). The tunneling between the dot and the
leads is described by $H_T$ with spin and orbital conservation since
the dot and the leads are usually fabricated within a same CNT and
thus have the same orbital symmetry \cite{Jarillo-Herrero1,
Jarillo-Herrero2, Makarovski1, Makarovski2}. Assuming some orbital
mixing, a crossover from $SU(4)$ to $SU(2)$ Kondo effect has been
investigated \cite{Choi,Lim}. Here we focus on a totally different
breaking of the $SU(4)$ Kondo effect by the spin-orbit interaction
which intrinsically lifts out the degeneracy in the single-particle
energy.

\begin{figure}[ht]
\includegraphics[width=0.8\columnwidth]{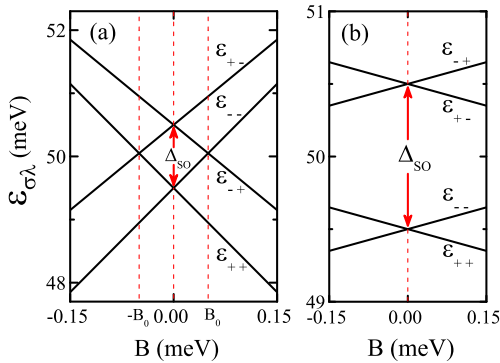}
\caption{$\varepsilon_{\sigma\lambda}$ as a function of (a) parallel
and (b) perpendicular magnetic fields for
$\varepsilon_d=50\textrm{meV}$, $\mu=10$, and
$\Delta_{SO}=1\textrm{meV}$.} \label{fig1}
\end{figure}

The energy spectrum of CNT QDs in low-energy limit is
$\varepsilon_{k}\simeq \hbar v_F\left\vert k\right\vert$ with $v_F$
the Fermi velocity and $k$ the quantized wave vector perpendicular
to the CNT axis. For a given wave vector $k_0>0$, $k=-\lambda k_0$
($\lambda=\pm$) denotes two degenerate graphene subbands,
corresponding to clockwise and anticlockwise classical orbits
encircling the tube circumference. In curved graphene, the
spin-orbit interaction is generally classified into three types
\cite{Huertas-Hernando2006}, an intrinsic coupling, a Rashba
coupling, and a curvature coupling. However, in most experimentally
accessible CNT QDs, only the last one is dominant
\cite{Huertas-Hernando2006}. It can be described by a spin-dependent
topological flux $\sigma\phi_{SO}$ and shifts $k$ by $\delta_1
k=\sigma\phi_{SO}/(r\phi_0)$ \cite{Huertas-Hernando2006,Kuemmeth},
with $\phi_0$ being the flux quantum and $r$ the tube radius. A
magnetic field $\mathbb{B}$ introduces an Aharonov-Bohm flux
$\phi_{AB}$ and further shifts $k$ by $\delta_2
k=\phi_{AB}/(r\phi_0)$, ending up with $k=-\lambda
k_0+\delta_1k+\delta_2k$ and thus $\left\vert
k\right\vert=k_0-\lambda\delta_1k-\lambda\delta_2k$. Combining
$\left\vert k\right\vert$ with the linear dispersion, and including
the spin Zeeman energy, the single-particle energy of the state
$\vert\sigma,\lambda\rangle$ reads $\varepsilon_{\sigma\lambda} =
\varepsilon_d-\sigma\lambda\Delta_{SO}/2-\lambda \mu
B\cos\theta-\sigma B$. Here $\varepsilon_d=\hbar v_Fk_0$ is the
basic dot level which depends on the confining geometry and can be
tuned by a gate voltage. The second term accounts for the spin-orbit
coupling with $\Delta_{SO}=2\hbar v_F\phi_{SO}/(r\phi_0)$. The last
two terms represent the orbital and spin Zeeman effects, where
$\mu=2\mu_{orb}/(g\mu_B)$ is the ratio between the orbital moment
$\mu_{orb}$ and the Bohr magneton $\mu_B$, the former is typically
one order of magnitude larger than the latter \cite{Minot},
$B=g\mu_B\mathbb{B}/2$ is the renormalized field, and $\theta$ is
the angle between the field and the CNT axis. Typical energy spectra
are shown in Fig.\,\ref{fig1}. Notably, the spin-orbit coupling
induces a zero-field splitting $\Delta_{SO}$ and two level crossings
for the parallel fields at $B=\pm B_0$ with
$B_0=\Delta_{SO}/(2\mu)$.

Electronic transport through the CNT QD is determined by the dot
retarded Green's function, which is
$G_m(\varepsilon)\equiv\langle\langle
d_{m}|d_{m}^{\dagger}\rangle\rangle=G_{m}^{0}(
\varepsilon)(1+U\sum_{m^{\prime}}\langle\langle
\hat{n}_{m^{\prime}}d_{m}|d_{m}^{\dagger}\rangle\rangle)$, where
$G_{m}^{0}\left(\varepsilon\right)=\left[\varepsilon
-\varepsilon_{m}-\Sigma^{0}\left(\varepsilon\right)\right]^{-1}$,
$\Sigma^{0}\left( \varepsilon\right)=\frac{\Gamma}{\pi}\int
\frac{\textrm{d}\varepsilon^\prime}{\varepsilon-\varepsilon^\prime}$,
$\Gamma=\sum_\alpha\Gamma_\alpha$, and $\Gamma_\alpha=\pi\rho\vert
V_\alpha\vert^2$ with $\rho$ being the density of states in the
leads at the Fermi level. Due to the interaction $U$,
$G_m(\varepsilon)$ includes a high-order Green's function, whose
equation of motion (EOM) produces in turn more higher-order ones. We
truncate this hierarchy by the Lacroix's decoupling procedure
\cite{Luo, Entin-Wohlman, Shiau} which successfully captures the
essential physics of the Kondo effect. By this scheme, in the limit
of $U\to\infty$, the dot Green's function is given by
\begin{equation}
G_{m}\left(\varepsilon\right) =\frac{1-\sum'_{m^{\prime}}\langle
\hat{n}_{m^{\prime}}\rangle -\sum'_{m^{\prime}}A_{m^{\prime}m}
}{[G_{m}^{0}(\varepsilon)]^{-1}+\sum'_{m^{\prime}}[\Sigma^{0}\left(\varepsilon\right)
A_{m^{\prime}m} -B_{m^{\prime}m}]}, \label{gd1}
\end{equation}
where the prime in the summation means $m'\neq m$ and
\begin{eqnarray}
&& \langle\hat{n}_{m^{\prime}}\rangle=-\frac{1}{\pi}\int
\textrm{d}\varepsilon f_{0}\left( \varepsilon\right)
\operatorname{Im}\left[ G_{m^{\prime}}\left( \varepsilon\right)
\right],\label{nd0}\\
&& A_{m^{\prime}m}=\frac{\Gamma}{\pi}\int
\textrm{d}\varepsilon^{\prime}f_{0}\left( \varepsilon^{\prime
}\right)\frac{\left[ G_{m^\prime }\left( \varepsilon
^{\prime}\right)  \right]
^{\ast}}{\varepsilon+\varepsilon_{m^{\prime}
}-\varepsilon_{m}-\varepsilon^{\prime}},\label{gda}\\
&& B_{m^{\prime}m}=\frac{\Gamma}{\pi}\int
\textrm{d}\varepsilon^{\prime}f_{0}\left(
\varepsilon^{\prime}\right)\frac{ 1+\left[
\Sigma^{0}\left(\varepsilon^\prime\right)G_{m^{\prime}}(
\varepsilon^{\prime}) \right] ^{\ast }
}{\varepsilon+\varepsilon_{m^{\prime}}-\varepsilon_{m}
-\varepsilon^{\prime}}, \label{gdb}
\end{eqnarray}
with $f_{0}\left(\varepsilon\right)
=\left(1/\Gamma\right)\sum_{\alpha} \Gamma_{\alpha}f_{\alpha}(
\varepsilon)$ where $f_{\alpha}( \varepsilon)$ represents the Fermi
distribution function in the leads. Eqs.\,(\ref{gd1})-(\ref{gdb})
can be solved self-consistently. For a QD with an $N$-fold
degenerate level, $\varepsilon_m=\varepsilon_d$, an $SU(N)$ Kondo
temperature within EOM is $T^{(N)}_K =
D\exp\{\pi\varepsilon_d/[(N-1)\Gamma]\}$ \cite{note}, where $D$ is
the half bandwidth in the leads and $\varepsilon_d$ is measured from
the Fermi level.

{\it Results and discussions.---}Experimentally, the measured
differential conductance $dI/dV$ near zero source-drain bias
directly reflects the Kondo features in the dot density of states
$\rho_d(\varepsilon)=-\left(1/\pi\right)\sum_m\textrm{Im}[G_m(\varepsilon)]$.
Within the Keldysh formalism \cite{book2}, the transport current is
$ I=\frac{4e}{h}\frac{\Gamma_L\Gamma_R}{\Gamma}\int
\textrm{d}\varepsilon\;[f_L(\varepsilon) -
f_R(\varepsilon)]\sum_m\textrm{Im}[G_m(\varepsilon)]$. In the
numerical results presented below, we use symmetric dot-lead
couplings $\Gamma_L=\Gamma_R$ and a symmetrically applied
source-drain bias $V$. The high-energy cutoff is fixed to be $D = 1$
and $\Gamma = 0.01D$.

\begin{figure}[ht]
\includegraphics[width=0.9\columnwidth]{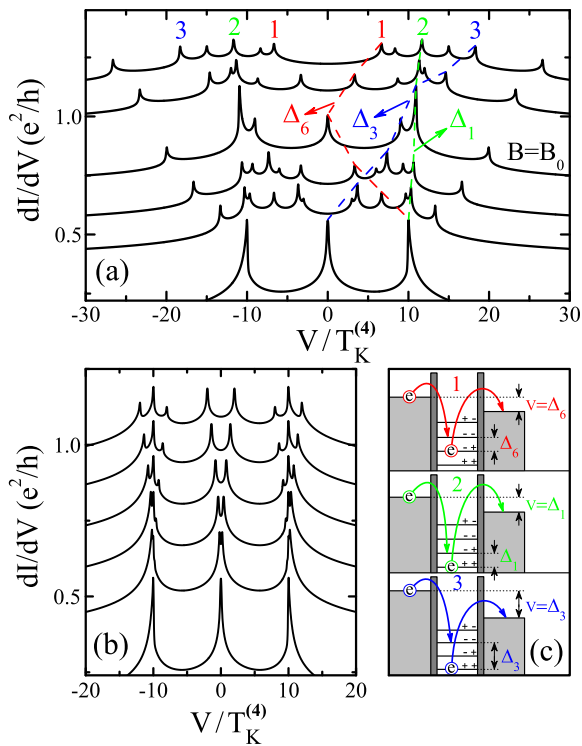}
\caption{Fine structures of the Kondo resonance with spin-orbit
coupling. (a) $dI/dV$ versus $V$ with various parallel magnetic
fields, $B/B_0=0$ (bottom), $1/3$, $2/3$, $1$, $4/3$, $5/3$ (top).
The red, green, and blue dashed lines are guides for the traces of
peak pairs marked by the numbers $1$, $2$, and $3$, respectively.
(b) $dI/dV$ versus $V$ with various perpendicular magnetic fields,
$B/T^{(4)}_K=0$ (bottom), $0.1$, $0.2$, $0.4$, $0.7$, $1$ (top). (c)
Schematic representation of three transition processes producing the
peak pairs $1$, $2$, and $3$ in (a), respectively. In (a) and (b),
the curves corresponding to $B\neq0$ are offset for clarity. The
parameters used are $\varepsilon_d=-10\Gamma$, $\mu=10$, and
$\Delta_{SO}=10T^{(4)}_K$.} \label{fig2}
\end{figure}

\begin{figure}[ht]
\includegraphics[width=0.8\columnwidth]{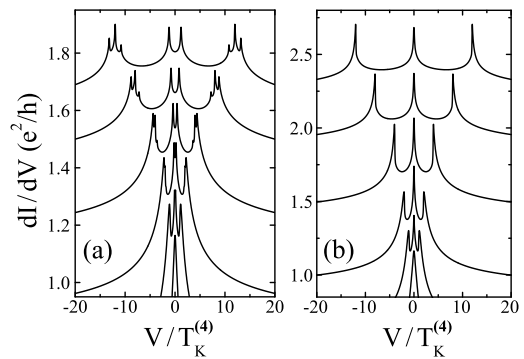}
\caption{$SU(4)$ Kondo splitting without spin-orbit coupling. (a)
$dI/dV$ versus $V$ with different parallel magnetic fields,
$B/T^{(4)}_K=0$ (bottom), $0.05$, $0.1$, $0.2$, $0.4$, $0.6$ (top).
(b) $dI/dV$ versus $V$ with different perpendicular magnetic fields,
$B/T^{(4)}_K=0$ (bottom), $0.5$, $1$, $2$, $4$, $6$ (top). The
curves corresponding to $B\neq0$ are offset for clarity. The
parameters used are $\varepsilon_d=-10\Gamma$, $\mu=10$, and
$\Delta_{SO}=0$.}\label{fig3}
\end{figure}

Figure \ref{fig2} presents our main results. Even at zero magnetic
field, the spin-orbit coupling lifts the degeneracy in parallel and
anti-parallel spin-orbit configurations of a single electron in the
dot [see Fig.\,\ref{fig1}], thereby breaking the $SU(4)$ symmetry of
the Kondo effect studied previously \cite{Choi, Lim, Busser,
Anders}. Instead, the Kondo effect manifests as three resonant peaks
in the differential conductance, which locate at $V=0$,
$\pm\Delta_{SO}$, respectively. Due to the coupling, each peak
entangles both spin and orbital degrees of freedom. Specifically,
the two side peaks arise from both spin-flip intraorbital and
spin-conserved interorbital transitions, while the central peak is
attributed to spin-flip interorbital transitions. A similar
zero-field three-peak structure has also been predicted in Silicon
QDs \cite{Shiau} but the underlying mechanism is entirely different.
In \cite{Shiau}, the two side peaks result from a trivial valley
(orbital) splitting at zero field and the remaining spin degeneracy
gives rise to the central peak.

When the magnetic field is applied, the Zeeman effect further
removes the remaining degeneracies in the energy spectrum and the
situation becomes complicated. In the field along the tube axis
[Fig.\,\ref{fig2}(a)], owing to the interplay of spin-orbit coupling
and Zeeman effect, each peak at zero field further splits into four
subpeaks, ending up with rich twelve-peak structures in the Kondo
resonance. At $B = B_0$, only seven peaks are visible and the other
peaks merge into the remaining ones. This is because of the level
degeneracy at this special field [see Fig.\,\ref{fig1}(a)]. Though
these fine multipeak structures appear to be complicated, the
inherent physics and the $B$-evolution of each peak can be clarified
by identifying all many-body cotunnelings that add up coherently to
screen both spin and orbital degrees of freedom.

In the one-electron regime, the Kondo effect arises from the
coherent superposition of six transition processes: two spin-flip
intraorbital transitions
$\vert\sigma,\lambda\rangle\Leftrightarrow\vert-\sigma,\lambda\rangle$,
two spin-flip interorbital transitions
$\vert\sigma,\lambda\rangle\Leftrightarrow\vert-\sigma,-\lambda\rangle$,
and two spin-conserved interorbital transitions
$\vert\sigma,\lambda\rangle\Leftrightarrow\vert\sigma,-\lambda\rangle$.
Each transition needs an energy of $\Delta_i$ ($i=1,\cdots,6$) and
develops a pair of Kondo peaks at $V=\pm\Delta_i$. From energy
differences between the initial and final states, one readily lists
all six transition energies as
$\Delta_1=\left\vert\varepsilon_{++}-\varepsilon_{-+}\right\vert$,
$\Delta_2=\left\vert\varepsilon_{+-}-\varepsilon_{--}\right\vert$,
$\Delta_3=\left\vert\varepsilon_{++}-\varepsilon_{--}\right\vert$,
$\Delta_4=\left\vert\varepsilon_{+-}-\varepsilon_{-+}\right\vert$,
$\Delta_5=\left\vert\varepsilon_{++}-\varepsilon_{+-}\right\vert$,
and
$\Delta_6=\left\vert\varepsilon_{-+}-\varepsilon_{--}\right\vert$.
Both the spin-orbit coupling and the magnetic field are
symmetry-breaking perturbations at the $SU(4)$ Kondo fixed point.
For $\Delta_{SO}=B=0$ and hence $\Delta_i=0$, the fixed point is
reached \cite{Choi,Lim}. Thus all the cotunneling processes are
elastic and constitute a highly symmetric $SU(4)$ Kondo resonance at
$V=0$ (see the curve corresponding to $B=0$ in Fig.\,\ref{fig3}). On
the other hand, one can uniquely identify the six transition
processes from the split Kondo peaks as long as all $\Delta_i$ are
different from each other and hence the six peak pairs (twelve
peaks) are well resolved, which requires $\Delta_{SO}\neq0$,
$B\neq0$, and $\cos\theta\neq0$. This is exactly the case of
Fig.\,\ref{fig2}(a). As an example, we trace the $B$-evolutions of
three peak pairs marked by the numbers $1$, $2$, and $3$ and
unambiguously attribute them to transition processes
$\vert-,+\rangle\Leftrightarrow\vert-,-\rangle$,
$\vert+,+\rangle\Leftrightarrow\vert-,+\rangle$, and
$\vert+,+\rangle\Leftrightarrow\vert-,-\rangle$, respectively, which
are schematically shown in Fig.\,\ref{fig2}(c). We also note that
such an unique identification is unavailable in Silicon QDs
\cite{Shiau} where no more than nine peaks are visible.

When the field is applied perpendicularly to the CNT, the orbital
Zeeman effect is absent and only the spin Zeeman effect is involved.
As shown in Fig.\,\ref{fig2}(b), while the central peak splits into
two subpeaks corresponding to the spin-flip interorbital
transitions, the two side peaks split into three subpeaks among
which the two split-off ones result from the spin-flip intraorbital
transitions and the rested one corresponds to the interorbital
transitions without spin-flip. In this case, some peaks are still
not resolved because $\Delta_3=\Delta_4$ and $\Delta_5=\Delta_6$.

It is useful to comment on the experimental observability of these
fine multipeak structures. On the one hand, in our EOM scheme the
Kondo temperature is underestimated and relatively, the splitting of
the Kondo peaks is more evident. On the other hand, the decoherence
\cite{Meir1993} neglected in the present study has an effect to
smear out the split peaks, especially those with high energies.
These two features will make the observation a bit difficult.
However, in ultra-clean CNT QDs and using highly resolved
spectroscopy measurements, it is still possible to observe part or
all of these fine structures.

For comparison, we plot in Fig.\,\ref{fig3} the multiple splitting
of the $SU(4)$ Kondo resonance by neglecting the spin-orbit
coupling. On increasing the field, the $SU(4)$ Kondo peak at the
zero bias splits in a simple way following the Zeeman effect, which
produces characteristic structures in agreement with the results in
the literature \cite{Jarillo-Herrero1, Choi, Lim, Makarovski1} but
quite different from that discussed above.

\begin{figure}[ht]
\includegraphics[width=0.8\columnwidth]{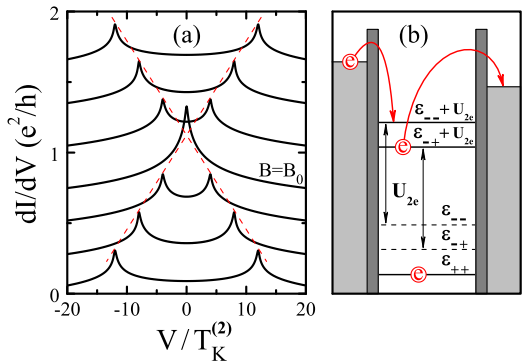}
\caption{Purely orbital Kondo effect in the $2e$ regime. (a) $dI/dV$
versus $V$ with various parallel magnetic fields,
$(B-B_0)/T^{(2)}_K=-0.6$ (bottom), $-0.4$, $-0.2$, $0$, $0.2$,
$0.4$, $0.6$ (top). The curves are offset for clarity and the
parameters used are $\varepsilon^\prime_{d}=-3\Gamma$ and $\mu=10$.
(b) Schematic diagram of the inherent many-body cotunneling.}
\label{fig4}
\end{figure}

The spin-orbit coupling further determines the filling order in the
$2e$ ground state \cite{Kuemmeth}, which exactly follows the first
excited state of the single-particle spectrum (Fig.\,\ref{fig1}). At
the parallel fields near the degenerate point $B_0$, while the first
electron always occupies the state $\vert +,+\rangle$, the second
electron can fluctuate between $\vert -,-\rangle$ and $\vert
-,+\rangle$, giving rise to a purely orbital Kondo effect as its
spin $\sigma=-$ is fixed. To describe this Kondo effect, $m$
appearing in all our previous formulae should be replaced by
$\{-,\lambda\}$ and the single-particle energy becomes
$\varepsilon^\prime_{-\lambda}\equiv\varepsilon_{-\lambda}+U_{2e}=\varepsilon^\prime_
{d}+(1-\lambda\mu)(B-B_0)$, where $U_{2e}$ is the $2e$ Coulomb
interaction and $\varepsilon^\prime_
{d}=\varepsilon_d+\Delta_{SO}/(2\mu)+U_{2e}$. Fig.\,\ref{fig4}(a)
presents the resulting $dI/dV$ as a function of $V$. For $B=B_0$,
there is a pronounced zero-bias peak which represents an $SU(2)$
orbital Kondo effect and provides a spin-polarized conducting
channel. The peak splits due to the field applied away from the
degenerate point. This orbital Kondo effect, as schematically shown
in Fig.\,\ref{fig4}(b), comes from the same shell and dwells in the
$2e$ valley, therefore being distinct from the one recently observed
in CNT QDs without spin-orbit coupling \cite{Jarillo-Herrero1}.

{\it Conclusion.---}We have studied the Kondo effect in a CNT QD
with spin-orbit coupling. It is shown that the Kondo effect
manifests as rich fine multipeak structures in the differential
conductance when a magnetic field is applied. These fine structures
are quite different from the $SU(4)$ Kondo effect studied previously
and might be observable in future experiments. In such a system, a
purely orbital Kondo effect develops in the $2e$ ground state due to
the particular multielectron filling order. Our results indicate
that the spin-orbit coupling significantly changes the low-energy
Kondo physics in CNT QDs.

Support from the NSFC (10575119), the Major State Basic Research
Developing Program (2007CB815004), and the Program for NCET of
China is acknowledged.


\begin{thebibliography}{99}
\bibitem{note}
This expression is slightly different from
$T^{(N)}_K\sim\exp[\pi\varepsilon_d/(N\Gamma)]$ given by the
renormalization group scaling theory \cite{Choi, Lim}. The
difference is due to the present truncation approximation, which can
be improved by involving more higher-order Green's functions
\cite{Luo}.
\end{thebibliography}
\end{document}